\documentclass[epj]{svjour}
\usepackage{bbm}
\usepackage{physics}
\usepackage{tikz,lipsum,lmodern}
\usepackage[export]{adjustbox}
\usepackage{amsmath}
\usepackage{amssymb}
\usepackage{xcolor}
\usepackage{graphicx}

\usepackage[unicode=true,pdfusetitle,
 bookmarks=true,bookmarksnumbered=false,bookmarksopen=false,
 breaklinks=false,pdfborder={0 0 0},backref=false,colorlinks=true,citecolor=blue,urlcolor=violet 
]{hyperref}
\newcommand{\orc}[1]{\href{https://orcid.org/#1}{\includegraphics[width=10pt]{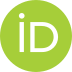}}}

\begin{document}
\title{A Generalized Formulation of Two-Particle Interference}

\author{Kamran Nazir\inst{1,2} \and Tabish Qureshi\inst{2}}
\institute{Department of Physics, Jamia Millia Islamia, New Delhi, India
\email{kamrannazir997@gmail.com}
\and
Centre for Theoretical Physics, Jamia Millia Islamia, New Delhi, India.
\email{tqureshi@jmi.ac.in}}

\abstract{
Two-photon interference is an interesting quantum phenomenon that is usually
captured in two distinct types of experiments, namely the Hanbury-Brown-Twiss
(HBT) experiment and the Hong-Ou-Mandel (HOM) experiment. While the HBT
experiment was carried out much earlier in 1956, with classical light, the 
demonstration of the HOM effect came much later in 1987. Unlike the former,
the latter has frequently been argued to be a purely quantum effect. A
generalized formulation of two-particle interference is presented here.
The HOM and the quantum HBT effects emerge as special cases in the 
general analysis. A realizable two-particle interference experiment, which
is intermediate between the two effects, is proposed and analyzed. Thus
two-particle interference is shown to be a single phenomenon with various
possible implementations, including the HBT and HOM setups.}

\maketitle

\section{Introduction}
Numerous fascinating phenomena, such as photon bunching and anti-bunching,
were seen with the development of lasers and quantum optics \cite{bunching}.
Whether they
are photons or neutral atoms, identical bosons in quantum optics have
been revealing an increasing number of fascinating aspects of quantum
mechanics \cite{aspect-qao}. 
Two-photon interference is a phenomenon which is at the heart of quantum
optics \cite{hbtlight1}. Two experiments which beautifully unveil
\emph{two-particle interference},
are the Hanbury-Brown-Twiss (HBT) experiment \cite{aspect-qao} and the
Hong-Ou-Mandel (HOM) experiment \cite{homrev}.

\begin{figure}[h]
\centering
\includegraphics[width=0.5\textwidth]{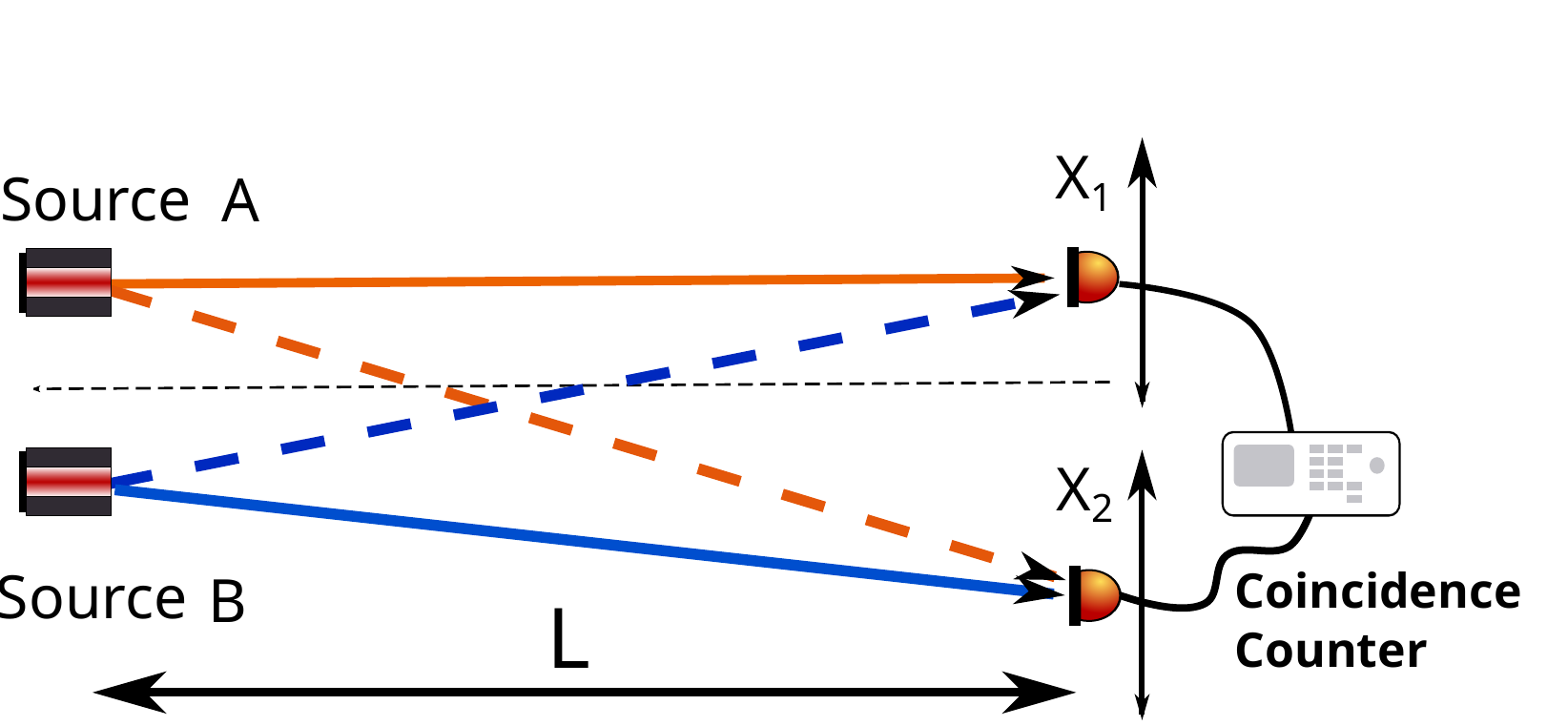}
\caption{A schematic diagram for the quantum Hanbury Brown-Twiss experiment.
Independent particles from sources A and B travel
and arrive at the two detectors at $x_1$ and $x_2$.}
\label{HBTexpt}
\end{figure}

\textbf{The Hanbury Brown-Twiss effect}

The HBT effect was discovered much before the HOM effect, in classical 
radio waves, and was used to measure the angular diameter of stars.
Later in 1956 it was demonstrated in classical light \cite{hbt}.
In the quantum version of HBT experiment \cite{aspect-qao}, two particles
emerge from two spatially separated 
sources A and B, and travel to separate, movable detectors at positions
$x_1$ and $x_2$ (see Fig. \ref{HBTexpt}). Individual detectors do not 
show any interesting effect, as expected. However, if one correlates the
\emph{intensity} of the two detectors, it shows an interference as 
function of the separation of the two detectors. This means that given
one photon has landed at a particular position, there are positions on which
the other photon would never land. This is quite an unexpected behavior for
independent photons.

The phenomenon can be understood easily using classical waves. However,
its applicability and meaning in quantum domain was widely debated and 
misunderstood \cite{hbtsilva}. People visualized that in order to show interference, the 
two photons, coming from independent sources, would need to ``know" where
to land! Now the HBT effect in the quantum domain is well understood
\cite{fano,tqhbt}. There is a crucial difference between the classical
and quantum HBT effect. For classical waves, the HBT interference visibility
can be at the most 1/2. However, in the quantum case the visibility can
ideally be 1. The HBT effect has now been demonstrated using ultracold 
atoms \cite{hbtatom1,hbtatom2,hbtatom3} and also with electrons
\cite{hbtelectrons}. Recenty a nonlocal HBT effect has also been 
demonstrated using entangled photons \cite{kiran}.

\textbf{The Hong-Ou-Mandel effect}

\begin{figure}[h]
\includegraphics[width=0.5\textwidth]{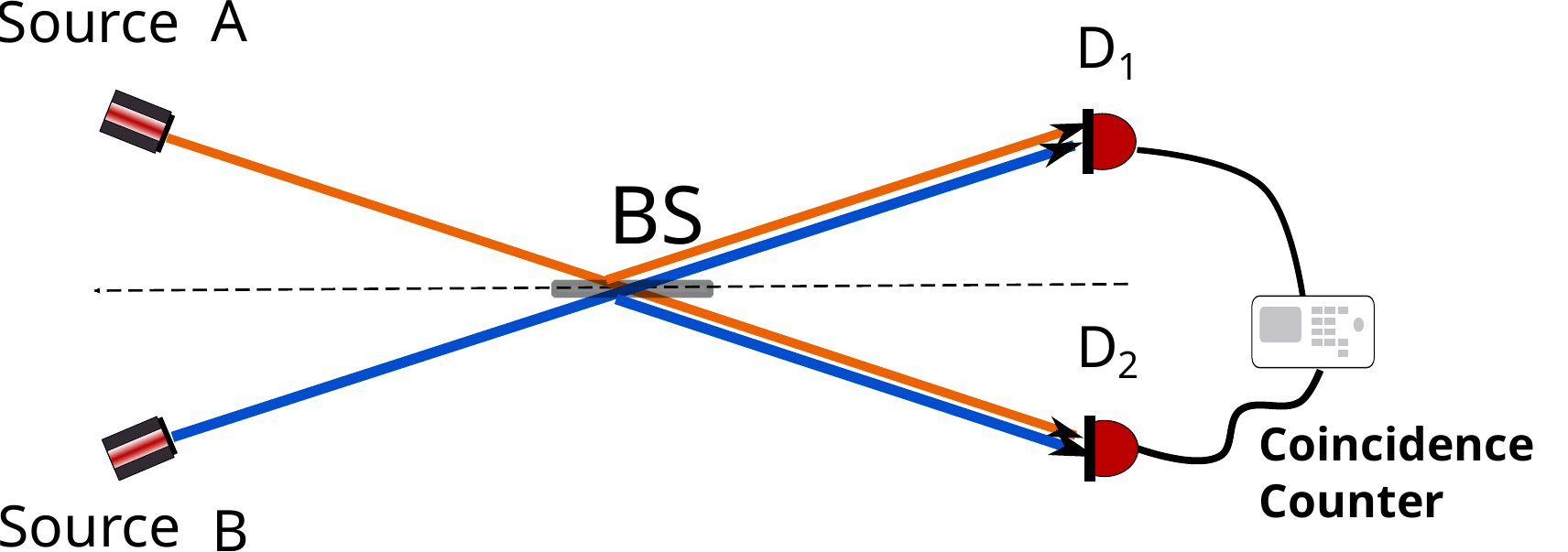}
\caption{A schematic diagram for the Hong-Ou-Mandel experiment.
Independent particles from sources A and B meet at the beam-splitter BS,
and then arrive at the detectors $D_1$ and $D_2$.}
\label{HOMexpt}
\end{figure}

We briefly introduce the HOM experiment which was first reported in
1987 \cite{hom}. Two identical particles emerge
from two spatially separated sources A and B (see Fig. \ref{HOMexpt}). 
The two particles are split by
the 50-50 beam-splitter BS, and reach the \emph{fixed} detectors $D_1,D_2$.
Since the two photons are independent, one would expect that half
the time the two would land up in different detectors. However, it is
observed that if the sources are tuned in such a way that the two photons
arrive at the beam splitter at the same time, the two photons always
go together to the same detector!
In other words, the coincident count of detectors $D_1$ and $D_2$
shows a dip, and goes to zero in the ideal case. This is the famous 
``HOM dip." The HOM effect has been demonstrated for completely independent
photons \cite{homlight2}.

In the way that the two effects have been described above, and also in
the way they came about historically, the two are quite distinct.
While HBT effect was originally seen in classical waves, the HOM effect is
believed to be completely quantum. However, it has now been demonstrated
that the HOM effect can also be realized with classical states
\cite{milman,urbasi}. There has been some work in which a
connection between the HBT and the HOM effects was established, in the
context of superradiant emission \cite{gsa1,gsa2}, in the context of a
$n$-photon generalized HBT effect \cite{zanthier} and in a Feynman
path-integral formulation \cite{liu}. It is clear that indistinguishability
of particles plays an important role in two-particle interference \cite{rps}.
Since at the quantum level both the effects are rooted in the
indistinguishability of identical particles,
one might wonder if the two can be formulated in a unified way. That is
the issue we address in this investigation, and demonstrate that indeed there
is a single two-particle interference phenomenon underlying both.

\section{Generalized $n$-port interferometer}

The connection between the HBT and HOM experiments can be understood by
drawing an analogy with the connection between a single particle two-slit
interference and the Mach-Zehnder interferometer \cite{tqtpd}. The
two-slit interference and the Mach-Zehnder interference are essentially
the same. The only difference is that while in the Mach-Zehnder interferometer,
a beam is split into two distinct beams, in the two-slit interference
experiment the beam emerging from one slit eventually spreads over an
infinite number of positions on the screen. We believe something similar
happens in the HOM and HBT experiments. Particles from one source are
split into two distinct beams by the beam-splitter in the HOM experiment.
On the other hand, particle emerging from a single source in the HBT 
experiment, spread over a continuous set of positions when they reach the
screen.

\begin{figure}[h]
\includegraphics[width=0.5\textwidth]{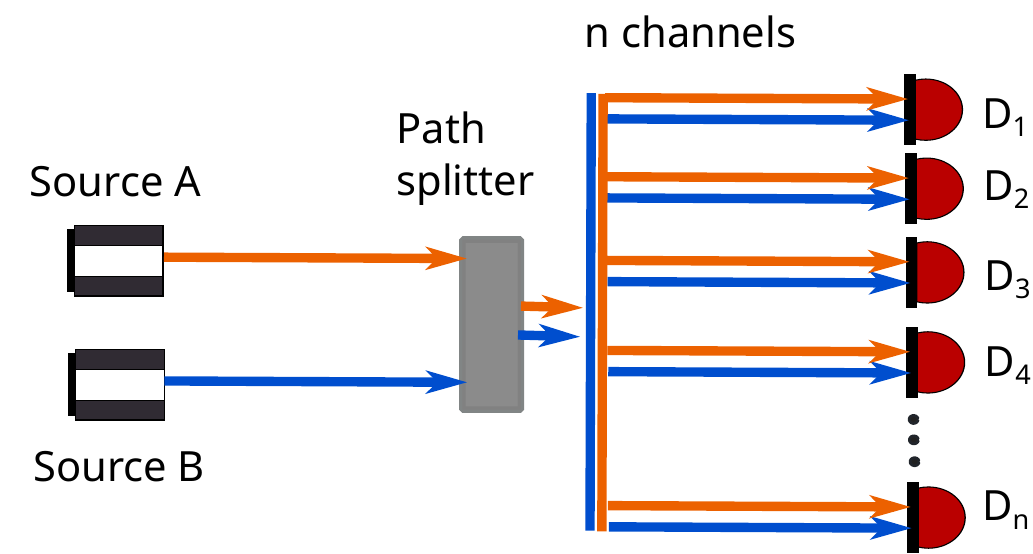}
\caption{Schematic diagram for a \emph{generalized} $n$-port two-particle
interference experiment. Independent particles from sources A and B are
split into $n$ common channels by the path-splitter, and then arrive at detectors
$D_1$ to $D_n$.}
\label{nport}
\end{figure}

In order to understand two-particle interference, we assume
a general scenario where there are two sources and a particle from a
particular source is split into $n$ channels. The same happens with the
particle coming from the other source. So the particles emerging from both
the sources go through the same channels, and can potentially arrive
at any of the $n$ detectors (see Fig. \ref{nport}). The two particles coming to
different channels may pick up different phases. Particle emanating from
source A has a state $|\psi_A\rangle$, and that from source B, a state
$|\psi_B\rangle$, such that the combined two particle state before
entering the path-splitter is given by a symmetrized product, on
account of them being identical bosons:
\begin{equation}
|\Psi_0\rangle = \tfrac{1}{\sqrt{2}}(|\psi_A\rangle_1|\psi_B\rangle_2
 + |\psi_A\rangle_2|\psi_B\rangle_1),
\label{Psi0}
\end{equation}
where the subscripts on the kets denote the particle label. Now each
particle gets split into an equal superposition of $n$ output channels.
Each channel $j$ ends up at unique detector $|D_j\rangle$. Thus the
effect of the path-splitter on the two initial states is given by
\begin{eqnarray}
\mathbf{U_{PS}}|\psi_A\rangle &=& \tfrac{1}{\sqrt{n}}\sum_{j=1}^n
e^{i\theta_j}|D_j\rangle\nonumber\\
\mathbf{U_{PS}}|\psi_B\rangle &=& \tfrac{1}{\sqrt{n}}\sum_{k=1}^n
e^{i\phi_k}|D_k\rangle,
\label{nsplit}
\end{eqnarray}
where $\theta_m, \phi_m$ are the phases picked up by the particles
in arriving in channel $m$, from source A and B, respectively.

The final two-particle state at the detectors is then given by
\begin{eqnarray}
|\Psi_f\rangle &=& \mathbf{U_{PS}}\tfrac{1}{\sqrt{2}}(|\psi_A\rangle_1|\psi_B\rangle_2
 + |\psi_A\rangle_2|\psi_B\rangle_1),\nonumber\\
&=& \tfrac{1}{n\sqrt{2}}\sum_{j=1}^n e^{i\theta_j}|D_j\rangle_1
\sum_{k=1}^n e^{i\phi_k}|D_k\rangle_2 \nonumber\\
&& + \tfrac{1}{n\sqrt{2}}\sum_{k=1}^n e^{i\theta_k}|D_k\rangle_2
\sum_{j=1}^n e^{i\phi_j}|D_j\rangle_1.
\end{eqnarray}
There are two kinds of terms in the product of the two sums. One is the
diagonal terms involving just one channel, and the other is the ``cross terms"
involving two channels. Latter ones potentially give rise to interference.
The final state then has the following form
\begin{eqnarray}
|\Psi_f\rangle &=& 
\tfrac{\sqrt{2}}{n}\sum_{j=1}^n e^{i(\theta_j+\phi_j)}|D_j\rangle_1
|D_j\rangle_2 \nonumber\\
&& + \tfrac{1}{n\sqrt{2}}\sum_{k\neq j}(e^{i(\theta_j+\phi_k)} +
e^{i(\theta_k+\phi_j)}) |D_j\rangle_1 |D_k\rangle_2 . \nonumber\\
\end{eqnarray}
In order to proceed any further we need some information on $n$ and the
various phases $\theta_j,\phi_k$.

\subsection{Arbitrary $n$: A simplified case}

For an arbitrary value of $n$, let us assume that all phases for the particle
coming from source A are zero, i.e., $\theta_j=0$ for $j=1,\dots,n$. For
particle coming from source B, we assume that $\phi_j=0$ for odd $j$,
and $\phi_j=\pi$ for even $j$. Now it is easy to see that the phase
factor in the cross-term
\begin{equation}
e^{i(\theta_j+\phi_k)} +
e^{i(\theta_k+\phi_j)} = \left\{
\begin{tabular}{lp{3cm}}
2 &~~($j,k$ both odd or both even)\\
0 &~~(in $j,k$ one is even, one odd)
\end{tabular} \right.
\end{equation}
\begin{figure}[h]
\includegraphics[width=0.5\textwidth]{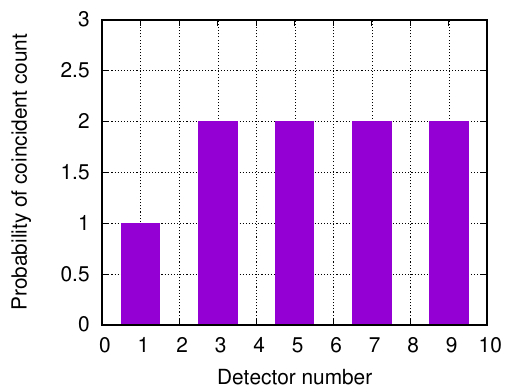}
\caption{Probability of coincident count of detector 1 with various
detectors, in units of $2/n^2$. The probability of coincident count with
even number detectors is zero. They constitute the dark fringes of the
interference pattern.}
\label{nccount}
\end{figure}
So the terms where one among $j,k$ is even, and the other odd, disappear.
This represents destructive interference. The detection results can be
summarized as follows. 
\begin{enumerate}
\item Probability of both particles landing at $j$'th detector
$= |_1\langle D_j|_2\langle D_j|\Psi_f\rangle|^2 = \frac{2}{n^2} $
\item Probability of particles landing at both odd or both even detectors 
$= |_1\langle D_j|_2\langle D_k|\Psi_f\rangle|^2 = \frac{4}{n^2} $. Such terms
represent the bright fringes, with two two-particle amplitudes adding up.
\item Probability of one particle landing at odd and one at even detector
$= |_1\langle D_j|_2\langle D_k|\Psi_f\rangle|^2 = 0 $. Such terms represent
the dark fringes, with two two-particle amplitudes destroying each other. 
\end{enumerate}
If one plots the probability of a coincident count of a particular detector
with various detectors, one would get an interference pattern, with
alternate detectors showing zero coincident count (see Fig. \ref{nccount}).
This general analysis can be used to study various real two-particle
interference experiments. We do that in the ensuing analysis.

\subsection{$n=2$: The HOM Experiment}

In the preceding analysis if we put $n=2$, it can exactly describe the HOM
experiment. In Fig. \ref{HOMexpt} if we assume that the \emph{lower} surface of
the mirror is half-silvered, then the photons coming from source A reach
the detectors $D_1, D_2$ without any phase change. However, the photons 
coming from source B pick up a phase of $\pi$ in reaching $D_2$. The
analysis for a general $n$ can then be applied here directly. We notice
that there are cross terms involving only one even and one odd detector.
Consequently in the coincident counts there will be only one dark fringe,
and no bright fringe. That is precisely what is seen in the HOM experiment.
The coincident count between the two detectors goes to zero.
\begin{figure}[h]
\includegraphics[width=0.5\textwidth]{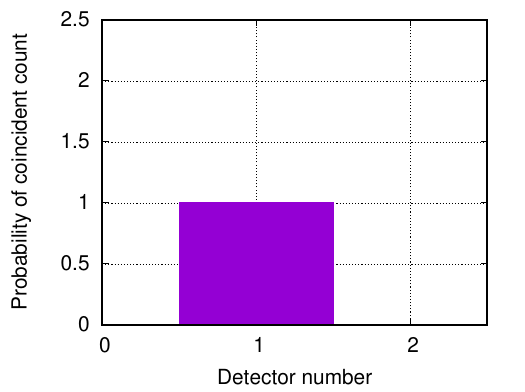}
\caption{Probability of coincident count of detector 1 with both the
detectors, in units of $1/2$. The probability of coincident count of
detector 1 with detector 2 is zero. The bar at detector 1 indicated that
both the particles land up at detector 1, half the time.}
\label{2ccount}
\end{figure}

\subsection{$n=4$: An Extended HOM Experiment}

\begin{figure}[h]
\includegraphics[width=0.5\textwidth]{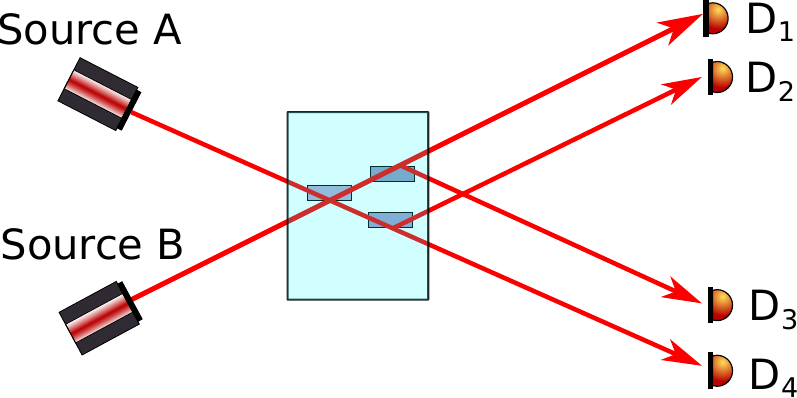}
\caption{Schematic diagram for a 4-port two-particle interference experiment.
Independent particles from sources A and B are
split into 4 common channels by a combination of 3 beam-splitters,
and then arrive at detectors $D_1, D_2, D_3, D_4$.}
\label{4port}
\end{figure}

In the HOM experiment, a particle from a particle source, is split by the
beam-splitter into a superposition of two parts, one reaching $D_1$ and
the other reaching $D_2$ (see Fig. \ref{HOMexpt}). Let us visualize an extended version of this
experiment where the particles coming from both the sources are split into a
superposition of four parts each. A realizable setup which implements
this scheme is shown in Fig. \ref{4port}.
\begin{figure}[h]
\includegraphics[width=0.5\textwidth]{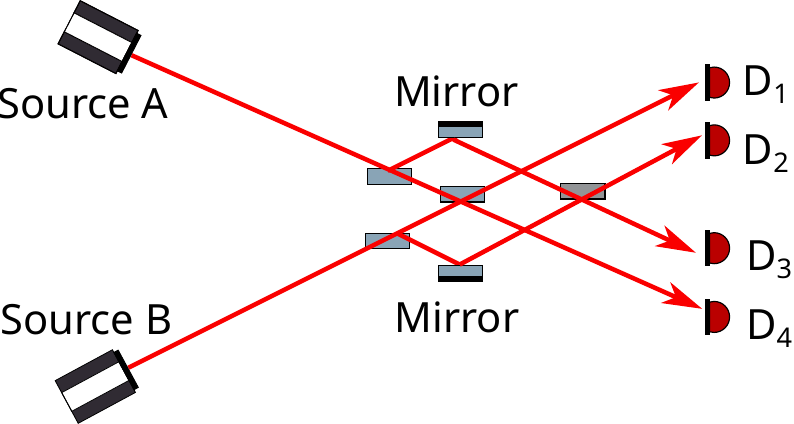}
\caption{Schematic diagram for an alternate 4-port two-particle interference
experiment. Independent particles from sources A and B are
split into 4 common channels by a combination of 4 beam-splitters,
and then arrive at detectors $D_1, D_2, D_3, D_4$.}
\label{4portnew}
\end{figure}
The combination of three beam-splitters plays the role of a 4-port 
path-splitter. The effect of the path-splitter on the particles coming
from source A and B, can be described as
\begin{eqnarray}
\mathbf{U_{PS}}|\psi_A\rangle &=& \tfrac{1}{2}(
|D_1\rangle+|D_2\rangle+|D_3\rangle+|D_4\rangle) \nonumber\\
\mathbf{U_{PS}}|\psi_B\rangle &=& \tfrac{1}{2}(
|D_1\rangle-|D_2\rangle+|D_3\rangle-|D_4\rangle) .
\end{eqnarray}
Now if the initial state before the two particles enter the path-splitter
is given by (\ref{Psi0}), the final state at the four detectors turns out
to be
\begin{eqnarray}
|\Psi_f\rangle &=& \mathbf{U_{PS}}\tfrac{1}{\sqrt{2}}(|\psi_A\rangle_1|\psi_B\rangle_2
 + |\psi_A\rangle_2|\psi_B\rangle_1),\nonumber\\
&=& \tfrac{\sqrt{2}}{4}\Big(|D_1\rangle_1|D_1\rangle_2 + |D_1\rangle_1|D_3\rangle_2
+ |D_2\rangle_1|D_2\rangle_2 \nonumber\\
&&+ |D_2\rangle_1|D_4\rangle_2 + |D_3\rangle_1|D_3\rangle_2
+ |D_3\rangle_1|D_1\rangle_2 \nonumber\\
&&+ |D_4\rangle_1|D_4\rangle_2 + |D_4\rangle_1|D_2\rangle_2 \Big) .
\end{eqnarray}
\begin{figure}[h]
\includegraphics[width=0.5\textwidth]{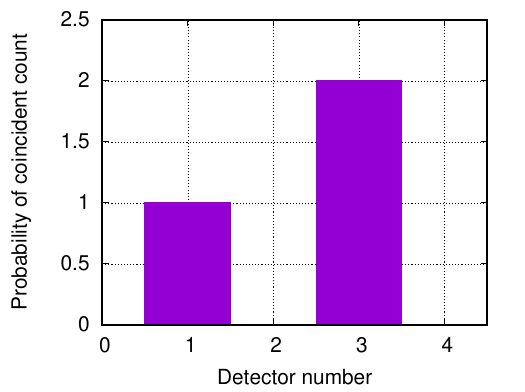}
\caption{Probability of coincident count of detector 1 with all four
detectors, in units of $1/8$.}
\label{4ccount}
\end{figure}
Notice that the terms $|D_1\rangle_1|D_3\rangle_2$ and
$|D_3\rangle_1|D_1\rangle_2$ both contribute to the same coincident count,
i.e., between detector 1 and 3. In the HOM experiment, the probability of
both the particles landing at a particular detector is 1/2, so the total
probability of both particles landing at $D_1$ or at $D_2$ add up to 1.
That is because there is no other possibility. However, in our extended
4-port HOM experiment, the probability of both particles landing at a
particular detector is not 1/4, rather it is 1/8. So the total probability
of both the particles landing up at the same detector doesn't add up to 1.
That is because there are other possibilities, e.g., of one particle going
to $D_1$ and the other to $D_3$

If one plots the probability of a coincident count of (say) $D_1$
with various detectors, one would get an interference pattern, with
two dark fringes (see Fig. \ref{4ccount}).
One would notice that although this case is an extension of the HOM
experiment, the result has some similarity with the HBT experiment where
one obtains an interference by doing coincident counts at detectors at
varying positions. Looking at Fig. \ref{4port}, some readers might get a
feeling that the first beam-splitter is already producing one bright channel,
and one dark channel, and it is trivial to see that a bright channel will be
split into two bright channels, and the dark channel into two dark channels,
by the two other beam-splitters. To address such an objection, a 4-port
interferometer can also be set up using an alternare arrangement
(see Fig. \ref{4portnew}), to which the preceding argument doesn't apply. 
However, the analysis remains identical to what has been presented here.

\subsection{$n\to\infty$: The HBT Experiment}

\begin{figure}[h]
\includegraphics[width=0.5\textwidth]{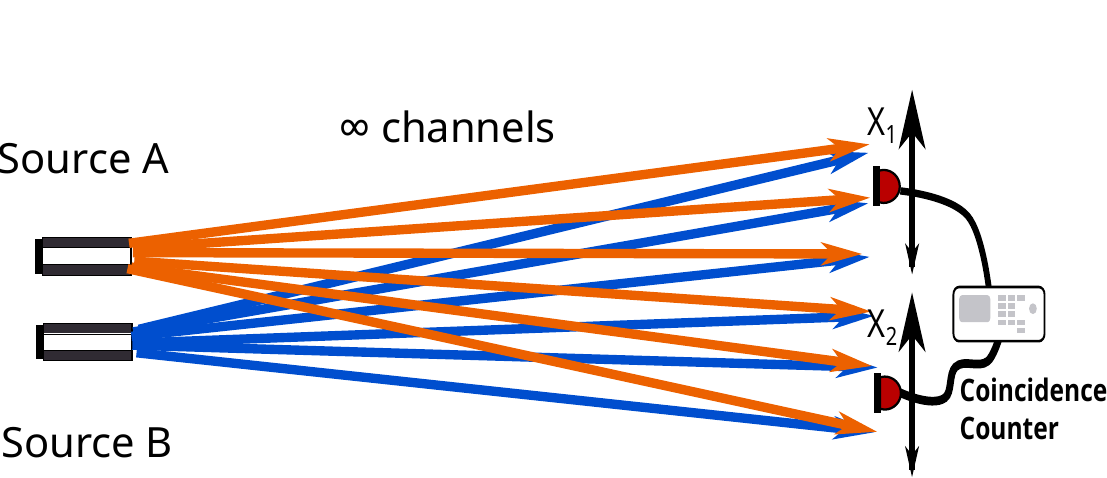}
\caption{Schematic diagram for the HBT experiment, as considered here.
A particle emerging a localized source gets split into a continuous number
of ``channels", by virtue of the expansion of the initially localized
wave-packet in space via Schr\"odinger evolution. On reaching the screen,
or the two movable detectors, the two wave-packets are delocalized over a
region of space, and strongly overlap with each other. The continuous 
positions are assumed to constitute an infinite number of channels.}
\label{hbt-schem}
\end{figure}

Next we investigate if the general $n$-port interferometer can capture the
HBT experiment. Here we consider a quantum HBT experimental setup, and not
the classical one. Here the particles are assumed to be emerging from
single-particle sources.
In the HBT experiment two particles emerge from two sources
A and B, localized at positions $x=\pm x_0$, respectively. The particles
travel along the y-axis and are finally detected at a continuous set of
positions $x_1$ and $x_2$ by two movable detectors (see Fig. \ref{hbt-schem}).
Essentially a particle emerging from a source is split into a continuous set
of infinite number of channels which end up at a continuous set of detector
positions.
There are no physical beam-splitters here, but the Schr\"odinger evolution
of the wave-packet of a particle, which is initially localized in space,
leads to its rapid expansion in the x-direction. On reaching the two
movable detectors, the wave-packet is strongly delocalized over a 
continuous set of positions along the x-axis. This continuous set of positions
can be considered as an infinite number of channels into which the particle
has been split.

The $n$-channel path-splitting described by (\ref{nsplit}) should
then be modified to take into account the continuous detector positions.
In order to normalize the probability in this continuous case, a position
dependent probability should be assigned to each channel, which is essentially
$|\psi(x)|^2dx$, $\psi(x)$ being the wavefunction of the particle in the
detection plane. The phases picked up by the channels are naturally position
dependent. The path-splitting can then be summarized as
\begin{eqnarray}
\mathbf{U_{PS}}|\psi_A\rangle &=& \int \psi(x) e^{i\theta_x}|x\rangle dx
\nonumber\\
\mathbf{U_{PS}}|\psi_B\rangle &=& \int \psi(x) e^{i\phi_x}|x\rangle dx,
\label{xsplit}
\end{eqnarray}
where $\theta_x, \phi_x$ are the phases picked up by the particles coming from
sources A and B, respectively, when they reach a position $x$ in the detection
plane. Here it has been assumed that $\psi(x)$ is approximately the same for
both the particles since $L \gg 2x_0$ (see Fig. \ref{HBTexpt}).
If $\lambda$ represents the real or de Broglie wavelength of a particle,
and it travels a distance $L$ along y-axis to reach the detector position $x$,
the phases acquired are given by $\theta_x = 2\pi x_0x/\lambda L$ and
$\phi_x = -2\pi x_0x/\lambda L$ \cite{tqhbt}. We do not specify the form of
$\psi(x)$ here - typically it is a Gaussian envelope. The final two-particle
state can then be written as
\begin{eqnarray}
|\Psi_f\rangle &=& \mathbf{U_{PS}}\tfrac{1}{\sqrt{2}}(|\psi_A\rangle_1|\psi_B\rangle_2
 + |\psi_A\rangle_2|\psi_B\rangle_1),\nonumber\\
&=& \tfrac{1}{\sqrt{2}}\int \psi(x) e^{i\theta_x}|x\rangle_1 dx
\int \psi(x') e^{i\phi_{x'}}|x'\rangle_2 dx' \nonumber\\
&& + \tfrac{1}{\sqrt{2}} \int \psi(x) e^{i\theta_x}|x\rangle_2 dx
\int \psi(x') e^{i\phi_{x'}}|x'\rangle_1 dx' .\nonumber\\
\end{eqnarray}
\begin{figure}[h]
\includegraphics[width=0.5\textwidth]{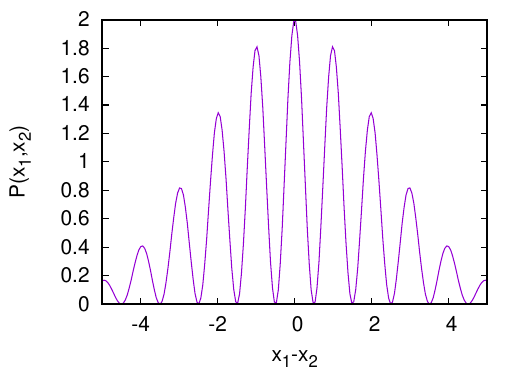}
\caption{Probability density (unnormalized) of coincident detection at
positions $x_1$ and $x_2$, against the detector separation $x_1-x_2$ (in the
units of the fringe width).
Here $x_1$ is fixed at 0 and $x_2$ is varied.}
\label{hbtplot}
\end{figure}
The probability amplitude of detecting one particle at $x_1$ and the other at $x_2$ is
then given by
\begin{eqnarray}
\Psi_f(x_1,x_2) &=& \tfrac{\psi(x_1)\psi(x_2)}{\sqrt{2}}\Big[
e^{\tfrac{i2\pi x_0(x_1-x_2)}{\lambda L}}\nonumber\\
&& + e^{\tfrac{-i2\pi x_0(x_1-x_2)}{\lambda L}}\Big] \nonumber\\
&=& \sqrt{2}\psi(x_1)\psi(x_2)
\cos\left(\tfrac{2\pi x_0(x_1-x_2)}{\lambda L}\right).\nonumber\\
\end{eqnarray}
The probability density of a coincident detection at positions $x_1, x_2$
is then given by
\begin{equation}
P(x_1,x_2) = |\psi(x_1)\psi(x_2)|^2\Big[ 1 + \cos\left(\tfrac{4\pi x_0(x_1-x_2)}{\lambda L}\right)\Big]
\label{hbtint}
\end{equation}
In expression (\ref{hbtint}) it is easy to see that there are values
of detector separation $x_1-x_2$ for which probability is zero. Those are
the dark fringes, and they are separated by $\lambda L/2x_0$. That is 
essentially the HBT effect. The probability density of a coincident detection
is plotted in Fig. \ref{hbtplot}, choosing $\psi(x)$ to be a Gaussian function.

\section{Discussion and Conclusion}
In the preceding section we formulated a general $n$-port two-particle
interferometer. It produces a generalized two-particle interference 
with $n$ detectors. For $n=2$ it reduces to the HOM experiment. For $n=4$ 
it represents an extended HOM experiment. This extended HOM experiment
can be realized without much difficulty. In the limit $n\to\infty$ when
the fixed detectors are replaced by two movable detectors in continuous space,
the generalized interferometer reduces to the HBT experiment.
Interestingly a multiport two-particle interferometer has very recently
been realized, in a somewhat different context \cite{sukhorukov}. 

Our generalized formulation reveals that two-particle interference is a single common
phenomena, with HOM and HBT experiments being its two specific cases,
among many possible ones. An earlier result showed that
a common duality relation exists, between the interference visibility and
particle distinguishability, for both HOM and HBT effects, indicating
a common origin for both \cite{tqtpd}. This duality relation was experimentally
confirmed too \cite{hong}. The present work demonstrates the equivalence of
the HBT and HOM effects in a more rigorous manner. It was earlier believed that
the HOM effect is a purely quantum effect whereas the HBT effect is possible
for classical waves too, although with a maximum visibility 1/2.
However,
it has now been demonstrated that the HOM effect can also be realized with
classical states \cite{milman,urbasi}.
So the final message is that
the two-particle interference should be viewed as a single
phenomenon with a variety of potential implementations, such as the HOM
and HBT configurations.

\section*{Declarations}
The authors have no conflicts to disclose.
There is no additional data associated with this manuscript.



\end{document}